\documentclass[prb,twocolumn,showpacs,amsmath,amssymb]{revtex4}
\usepackage{graphicx}
\usepackage{dcolumn}
\usepackage{bm}
\usepackage{epsfig}
\begin{document}

\title{Local magnetism and magnetoelectric effect in 
HoMnO$_{3}$ studied with muon-spin relaxation}

\author{H.J. Lewtas}
\author{T. Lancaster}
\author{P.J. Baker}
\author{S.J. Blundell}
\author{D. Prabhakaran}
\affiliation{
Oxford University Department of Physics, Clarendon Laboratory,  Parks
Road, Oxford, OX1 3PU, UK
}
\author{F.L. Pratt}
\affiliation{
ISIS Facility, Rutherford Appleton Laboratory, Chilton, Oxfordshire OX11 0QX, UK}

\date{\today}

\begin{abstract}
We present the results of muon-spin relaxation ($\mu^{+}$SR)
measurements on the hexagonal manganite HoMnO$_{3}$. 
Features in the temperature-dependent relaxation rate $\lambda$ correlate with the
magnetic transitions at 76~K, 38~K and 34~K. The highest
temperature transition, associated with the ordering of Mn$^{3+}$ moments has the largest effect on $\lambda$. 
The application of a static electric field of $E=10^{4}$~Vm$^{-1}$ 
below $T=50$~K causes a small reduction in $\lambda$
which is suggestive of coupling between ferroelectric and magnetic domain walls in the ordered state of
the material. 
\end{abstract}

\maketitle

\begin{figure}
\begin{center}
\epsfig{file=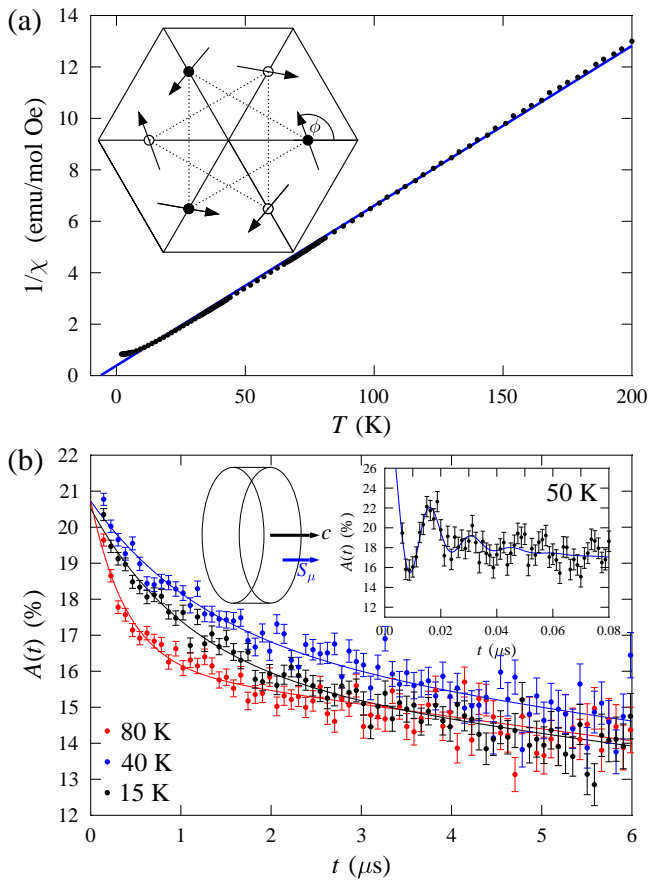,width=\columnwidth}
\caption{
(a) Evolution of the inverse field-cooled magnetic susceptibility of HoMnO$_{3}$ with temperature. 
{\it Inset}: magnetic order in 
in Mn$^{3+}$ spin system characterized by the angle $\phi$.
(b) ZF $\mu^{+}$SR data for HoMnO$_{3}$  measured with a pulsed muon source.
The relaxation is exponential across the entire
measured temperature regime. 
{\it Inset, left}: the experimental configuration with the initial
muon spin lying parallel to the crystallographic $c$-axis.  Filled circles represent atoms on the
$z=0$ plane, open circles in the $z=c/2$ plane.
{\it Inset, right}: ZF $\mu^{+}$SR data at $T=50$~K measured at early times using a continuous muon source.
Antiferromagnetic order gives rise to the oscillations in the data. 
\label{data}}
\end{center}
\end{figure}

\section{Introduction}

The current resurgence of interest in multiferroic materials is driven
by the possibility of controlling electrical charge using applied
magnetic fields and magnetic moments by applied voltages.  A number of
compounds have been identified as being of interest in this respect
\cite{cheongmostovoy}, and recent discoveries of large magnetoelectric
responses \cite{hill00,fiebig02,kimura03,hur04,lottermoser} have
reinvigorated the field \cite{fiebig,rameshspaldin}.  One such family
of compounds is the hexagonal manganites, $R$MnO$_3$ (R=Ho, Er, Tm, Yb,
Lu, Y), in which an electrical dipole moment results from a
high-temperature structural transition.  This arises because of a
nonlinear coupling to nonpolar lattice distortions associated with the
buckling of $R$--O planes and tilts of the MnO$_6$ bipyramids
\cite{vanaken}.  In particular, large magnetoelectric effects have
been found in hexagonal HoMnO$_{3}$ \cite{lottermoser,lorenz,hur09} and
further studies of this compound have ensued, motivated in part by its
rich phase diagram \cite{lorenz,vajk05,yen05,delacruz}.  Magnetoelectric
coupling occurs because superexchange interactions depend sensitively
on orbital overlap, and these can be tuned with an electric field as
metal cations and ligand anions move in opposite directions
\cite{lueken,gehring}.  Furthermore, the symmetry requirements of
magnetoelectric coupling are stringent and the lower symmetry inside
domain walls may allow magnetoelectric coupling between wall
magnetization and ferroelectric polarization \cite{domains}.

One method to attempt to observe such coupling at a microscopic level
is via muon-spin relaxation ($\mu^{+}$SR).  This technique is a sensitive
probe of the spin distributions in a magnetic material and has proven
particularly useful in probing frustration related effects\cite{bert}
including the case of the YMnO$_{3}$ \cite{lancaster07}. $\mu^{+}$SR can be
combined with electric fields, particularly at pulsed sources of muons
where the electric field can be switched on or off between muon pulses,
allowing even small effects to be measured.  Although combined
electric field and $\mu^{+}$SR techniques have been extensively employed
in the study of electronic states in semiconductors \cite{eschenko},
it has only recently been applied to magnetic systems.\cite{storchak}

In this paper, we describe the results of $\mu^{+}$SR measurements on
hexagonal HoMnO$_3$ as a function of
temperature and also applied electric field.  We identify the source
of the observed spin relaxation, follow this through the various
magnetic transitions observed in the absence of an applied field, and
show that a small, but measurable, contribution to the relaxation can
be adjusted using an applied electric field, an effect we associate
with the behavior at domain walls.  This paper is structured as
follows: in section II we review the magnetic and electric properties
of hexagonal HoMnO$_3$, in section III we describe the experiments and
in section IV we present the results and discussion.

\section{Properties of $\mathbf{HoMnO}_{3}$}

\subsection{Magnetic properties}

The magnetic phase diagram of HoMnO$_{3}$ has been the subject of much experimental investigation
owing to the existence of a large number of magnetic field-dependent phases
\cite{munoz,lonkai,fiebig02,vajk05,yen}.
The magnetic system is based around triangular layers of $S=2$ Mn$^{3+}$ spins with
Ho$^{3+}$ spins between layers.
The Mn$^{3+}$ spins are confined by single-ion anisotropy to the $a$-$b$ plane, and 
are magnetically frustrated due to the triangular geometry and antiferromagnetic (AFM)
coupling. 
The Ho$^{3+}$ spins, on the other hand, possess an Ising-like anisotropy forcing them to align
parallel or antiparallel to the $c$-axis.
The Mn$^{3+}$ sublattice relieves its frustration via 
ordering in the $a$--$b$ plane below $T_{\mathrm{N}} \approx 72$~K, where the Mn$^{3+}$ spins 
adopt a 120$^{\circ}$ structure of the P6$_{3}$ type, shown inset in Fig.~\ref{data}(a).
As the temperature is reduced in zero applied magnetic field, several magnetic 
phases are realized which are distinguished by  the absence or presence of Ho$^{3+}$ ordering and
the angle $\phi$ which the Mn$^{3+}$ spins
make to the local $x$ direction (Fig.~\ref{data}(a)).
This angle differs by $\pi/2$ in each of the ordered phases 
(with the exception of a small intermediate phase described below).

The HTI phase (P6$^{\prime}_{3}$c$^{\prime}$m) occurs for $38$~K$\lesssim T <T_{\mathrm{N}}$ where $\phi = \frac{\pi}{2}$. 
In this phase the Ho$^{3+}$ spins are disordered.
Below $T \approx 38$~K the Ho$^{3+}$ moments (at least partially) order antiferromagnetically 
in the $c$ direction 
\cite{nandi08} while Mn spin reorientations take place.
In the small temperature region $34 \lesssim T \lesssim 38$~K there is evidence of 
an intermediate magnetic phase (IP) (space group P6$^{\prime}_{3}$) 
characterized by an angle taking values\cite{lorenz} $ \frac{\pi}{2} < \phi < \pi$.
On cooling below $T \approx 34$~K, $\phi$ locks in to $\phi=\pi$ 
(the HTII phase, P6$^{\prime}_{3}$cm$^{\prime}$).
Finally, a further Mn$^{3+}$ spin rotation transition occurs around 5~K to the low temperature (LT) phase where 
$\phi=\frac{3 \pi}{2}$ (P6$_{3}$cm). Here the Ho$^{3+}$ moments also reorder to another structure, although 
they are still polarized along the $c$-direction. 

\subsection{Electric properties}

HoMnO$_{3}$ is a high $T_{\mathrm{C}}$ ferroelectric ($T_{\mathrm{C}}=875$~K) with very large polarization 
of $P \sim 56$~mCm$^{-2}$ directed along the $c$-axis \cite{coeure96} caused by vertical Ho--O displacements.
There is also a strong magnetoelectric (ME) effect, seen as a change in electric polarization larger than 
$\delta P \sim 80$~ $\mu$C m$^{-2}$ in a modest magnetic field of a few tesla\cite{hur09}.

Measurements of optical second harmonic generation (SHG) and Faraday rotation suggest that 
the application of an electric field (E-field) of $E=10^{5}$V m$^{-1}$ changes the local 
magnetic properties of the system considerably.
At all temperatures below $T_{\mathrm{N}}$ the SHG signal is quenched
in the presence of $E$. This was interpreted as resulting from the E-field changing
the magnetic structure of the Mn$^{3+}$ sublattice to a $\phi=0$ arrangement
and, in addition,
the Ho$^{3+}$ sublattice being forced to order ferromagnetically at all temperatures 
below $T_{\mathrm{N}}$. This results in
a  magnetic structure described by the P6$_{3}$c$^{\prime}$m$^{\prime}$ spacegroup \cite{lottermoser}.
However, this picture of an E-field induced ferromagnetic Ho$^{3+}$ sublattice
contrasts with more recent magnetic x-ray results, where it was concluded that the
antiferromagnetic Ho$^{3+}$ magnetic structure remains unchanged by the application 
fields $E \leq 10^{7}$~Vm$^{-1}$ in the region\cite{nandi08} $5< T < 39$~K. It has also been 
pointed out\cite{lorentz04} that FE and AFM domain structure may be important in 
interpreting magnetoelectric effects in this material.

\section{Experimental details}

Samples of HoMnO$_{3}$ were prepared as described previously \cite{zhou05}.
Stoichiometric amount of dried Ho$_{2}$O$_{3}$ and MnO$_{2}$ powders were mixed and
calcined in air for 24 hours at 1100~$^{\circ}$C and 1200~$^{\circ}$C respectively with
intermediate grinding. The powders were formed into
cylindrical feed rods and sintered at 1250~$^{\circ}$C for 15 hours in air. Single
crystal of HoMnO$_{3}$ were grown using the floating-zone technique. 
The growth was carried out at a rate of 3~mm h$^{-1}$ under the flow of mixed Ar/O$_{2}$ gas.
Crystals were characterized using x-ray diffraction and dc magnetic susceptibility measurements.

In a $\mu^{+}$SR experiment, spin-polarized
positive muons are stopped in a target sample, where the muon usually
occupies an interstitial position in the crystal.
The observed property in the experiment is the time evolution of the
muon spin polarization, the behavior of which depends on the
local magnetic field $B$ at
the muon site, and which is proportional to the
positron asymmetry function \cite{steve} $A(t)$.
ZF $\mu^{+}$SR  measurements were made using the 
GPS spectrometer at the Swiss Muon Source (S$\mu$S) and 
the EMU spectrometer at the ISIS facility. 
E-field $\mu^{+}$SR measurements were also made using the EMU spectrometer.
For the measurements at S$\mu$S an unaligned polycrystalline sample was measured,
while for the measurements at ISIS a mosaic of aligned crystals was prepared
in order that the E-field could be applied along a single crystallographic direction. 
To produce the mosaic, HoMnO$_{3}$ crystals were polished into thin
plates of thickness 1~mm  with the $a$--$b$ plane forming the polished surfaces. 
Crystallites were aligned such that the muon-spin was initially parallel to the
crystallographic $c$-direction. 
Gold electrodes of thickness 1~$\mu$m were evaporated onto these surfaces 
allowing the application of E-fields along $c$.
The sample were masked with insulating PTFE and mounted in a aluminum sample holder. 

\section{Results and discussion}

The dc magnetic susceptibility (Fig.~\ref{data}(a)) shows no feature that indicates the
onset of magnetic ordering at 76~K. This is consistent with previous measurements of the susceptibility
\cite{pauthenet} and is likely to be caused by the high paramagnetic
susceptibility of the Ho$^{3+}$ ions hiding the magnetic transition. No
difference between the zero-field-cooled and field-cooled data is observed at low temperatures. 

\subsection{ZF $\mu^{+}$SR}

Fig.~\ref{data}(b) shows example ZF $\mu^{+}$SR spectra measured on HoMnO$_{3}$. 
The inset shows an example spectrum  
measured on the unaligned polycrystalline sample at S$\mu$S at $T=50$~K.  Spontaneous oscillations
are seen at early times ($0 \leq t \lesssim 0.1$~$\mu$s) in all spectra measured below $T_{\mathrm{N}}$.
Upon cooling to the HTI phase the spectra consist of oscillations at a single frequency with a large relaxation rate. In the
HTII phase the oscillations change their form and there is evidence for a second frequency component. 
These observations are in good agreement with the results of previous $\mu^{+}$SR measurements\cite{barsov07} 
made on powder samples of HoMnO$_{3}$, where oscillations were also observed below $T_{\mathrm{N}}$. 

In contrast to the measurements made at S$\mu$S, those 
made on a single crystal using the pulsed muon source at ISIS did not show resolvable oscillations.
This is because the pulse length $\tau_{\mathrm{pulse}}$ limits the dynamic range of the measurement 
to rates $\sim 1/\tau_{\mathrm{pulse}}$. For muons with their initial spin direction oriented at an angle $\theta$
to a static magnetic field, we expect a spectrum described (in the absence of fluctuations) by
\begin{equation}
A(t) = A_{0}\left[ \cos^{2} \theta + \sin^{2}\theta \cos(\gamma_{\mu} B t) \right],
\label{angles}
\end{equation}
where $\gamma_{\mu}$ is the muon gyromagnetic ratio and $B$ is the local magnetic field at the muon site. 
When oscillations are not resolvable, only the first term in Eq.~(\ref{angles}) is measured, 
giving a signal whose amplitude is due to that component of the muon spin initially oriented parallel 
to the local magnetic field at the muon site. The relaxation of such a signal is due to the dynamic
fluctuations of the local field at the muon site. 
In the fast magnetic fluctuation regime (typical of most magnetic materials)
dynamics give rise to exponential relaxation\cite{hayano}
$A(t) = A_{0} \exp (-\lambda t) \cos^{2}\theta $
with a relaxation
rate $\lambda$ given by $\lambda = 2 \gamma_{\mu}^{2} \langle B^{2} \rangle \tau$, where 
$\langle B^{2} \rangle$ is the second moment of the magnetic field distribution at the muon sites and $\tau$ is 
the correlation time describing the dynamics of the local field distribution. 
Dynamic fluctuation in the paramagnetic phase may also be expected to give rise to exponential relaxation, although
in that case there is no angle dependence and $A(t)=A_{0} \exp(-\lambda t)$.
The spectra measured at ISIS show this expected exponential relaxation at all measured temperatures 
($15< T < 300$~K). The measured asymmetry for these data is best described by a fitting function
\begin{equation}
A(t) = A_{1} e^{-\lambda t} + A_{\mathrm{bg}},
\label{fiteq}
\end{equation}
where $A_{\mathrm{bg}}$ represents the contribution from those muons that stop in the sample
holder or cryostat tail. Note that the background signal due to muons stopping in PTFE is the distinctive
F--$\mu^{+}$--F state \cite{lancaster09fmuf}, so can be easily identified and subtracted.

\begin{figure}
\begin{center}
\epsfig{file=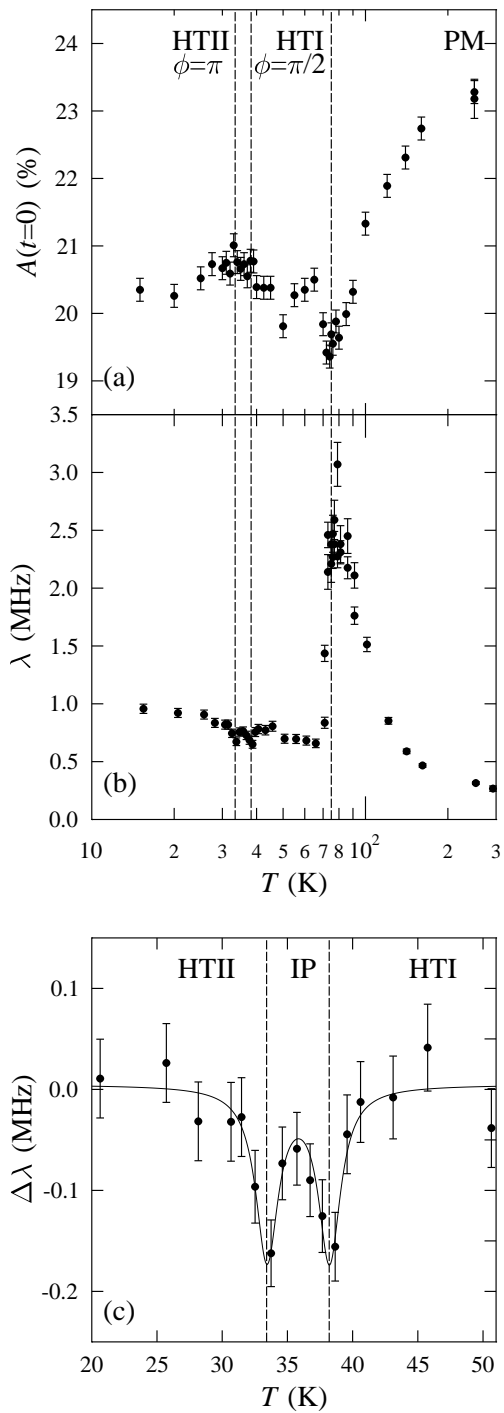,width=7cm}
\caption{
Results for measurements made in ZF across the phase diagram.
(a) Temperature dependence of the initial asymmetry $A(t=0)$.
(b) Evolution of the relaxation rate $\lambda$ with temperature.
(c) Detail from (b) after background subtraction showing the behavior of $\lambda$ at low temperature. 
Minima are
observed at the phase boundary between HTI and IP phases and at that between IP and HTII
phases.
\label{lambda}}
\end{center}
\end{figure}

Although it might be hoped that the temperature evolution of $A(t=0)=A_{1}$ in Eq.~(\ref{fiteq}) would tell us the
orientation of the local magnetic field at the muon site via Eq.~(\ref{angles}), this is not the case here. 
This quantity (Fig.~\ref{lambda}(a)) shows a decrease in the paramagnetic regime as temperature is reduced 
towards $T_{\mathrm{N}}$.
This points to the presence of an additional fast relaxation component that we do not resolve. The presence of
two relaxation rates signals either two classes of muon site or to two relaxation channels (see 
Ref.~\onlinecite{lancaster09ii} for a discussion). 
The amplitude $A(t)=0$ then shows quite scattered behavior in the HTI phase
before levelling off in the HTII phase. The presence of an additional relaxing component makes any 
determination of the local field direction in the crystal very difficult.

In contrast, the relaxation rate $\lambda$ provides a good probe of the magnetic behavior in HoMnO$_{3}$.
On cooling from $T=300$~K the rate $\lambda$ (Fig.~\ref{lambda}(b)) 
is seen to increase with decreasing temperature,
peaking around 79~K. This reflects the slowing of fluctuations of the Mn$^{3+}$ moments as the material 
approaches the magnetic
phase transition at $T_{\mathrm{N}}$ where these moments magnetically order. 
On cooling further we observe that with decreasing temperature $\lambda$
shows a weak, approximately linear, increase. 
Subtracting this trend from the data results in two
pronounced minima, shown in Fig.~\ref{lambda}(c), which occur at the two proposed transition temperatures 
between the phases HTI, IP and HTII. Fitting lorentzian lineshapes to these features
allows us to estimate the transition temperatures as $T_{\mathrm{IP}}=38.2(2)$~K and $T_{\mathrm{HTII}}=33.4(3)$~K. 
We note from our previous analysis that minima in $\lambda$ near the phase boundaries at $T_{\mathrm{IP}}$ and $T_{\mathrm{HTII}}$ would imply 
that $\langle B^{2} \rangle \tau$ goes through minima at these temperatures. This would
correspond to correlation times which are shortest at these phase boundaries.

We note here that the largest change in $\lambda$ is seen at $T=79$~K where the Mn$^{3+}$ spins first 
order and at which the Ho$^{3+}$ spins have do not undergo any change. 
This contrasts with the changes seen below 38~K, 
where the Ho$^{3+}$ spins begin to order, which are far more modest. 
This implies that, in our measurements,
we are mostly sensitive to the 
fields due to the local magnetic fields due to the Mn spins.
This is consistent with the case of structurally similar 
YMnO$_{3}$\cite{lancaster07}, where
ZF $\mu^{+}$SR measurements revealed the existence of two separate classes of muon
site, showing quite differently behaving relaxation rates. 
In YMnO$_{3}$, one site appeared to be closely coupled to the magnetic layers of Mn$^{3+}$ ions, 
while the other was 
less well defined, but was probably due to sites between the layers. 
We might expect the muon sites in HoMnO$_{3}$ to be similar, with the site near the 
Mn layers giving rise to the relaxation with rate $\lambda$ and the site between layers,
coupled more strongly to the Ho$^{3+}$ moments. This latter site would then be responsible
for rapid relaxation in the PM phase which leads to the loss of initial asymmetry described above.

\subsection{E-field $\mu^{+}$SR}
E-field $\mu^{+}$SR measurements were made in the presence of a longitudinal magnetic field (LF)
 $B_{\mathrm{a}}$
directed along the crystal's $c$-axis. 
Static electric fields of $E=10^{4}$~Vm$^{-1}$ were applied in the same direction. 
The saturation field of HoMnO$_{3}$ \cite{lottermoser} occurs at $E=10^{4}$~Vm$^{-1}$, above which we expect
a single ferroelectric domain.

\begin{figure}
\begin{center}
\epsfig{file=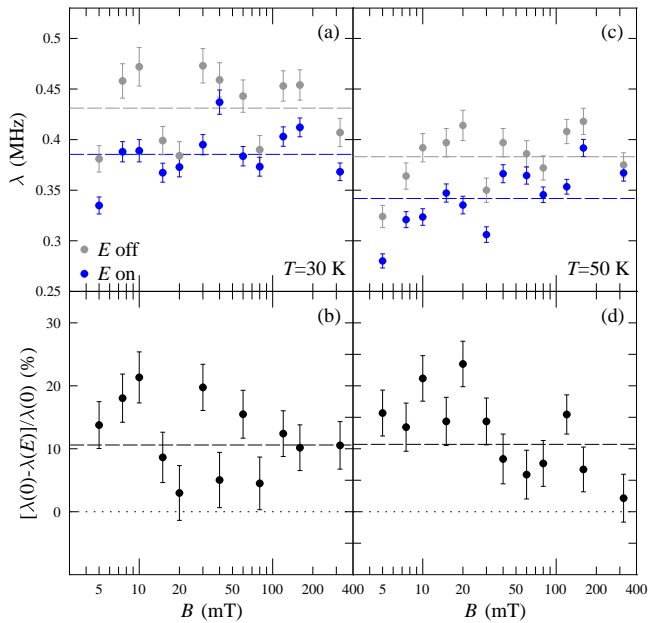,width=\columnwidth}
\caption{
(a) Relaxation rate $\lambda$ at $T=30$~K (HTII phase) in the presence and absence of an
electric field. 
The average value for each condition is shown as a dashed
line. 
The field is seen to reduce the mean value of $\lambda$.
(b) Fractional difference in $\lambda$ defined as 
$[\lambda(E=0)-\lambda(E)]/\lambda(E=0)$. The average difference is around 
10\% (dashed lines).
(c) and (d) same as (a) and (b) but measured at $T=50$~K (HTI phase).
\label{efield}}
\end{center}
\end{figure}

As in the ZF case, the spectra for E-field $\mu^{+}$SR measurements showed exponential relaxation described by
relaxation rate $\lambda$. Fig.~\ref{efield}(a) and (c) show the results of applying a magnetic field
$B_{a}$ in the presence and absence of a static electric field  $E$ in the HTII
phase (at $T=30$~K) and the HTI phase (at $T=50$~K).
In both phases there is little systematic variation of $\lambda$ as a function of applied magnetic field 
$B_{\mathrm{a}}$,
independent of the presence (or absence) of the $E$-field. It is expected that a dynamic relaxation rate 
should vary with applied  magnetic field as
$
\lambda = 2 \gamma^{2}_{\mu} \langle B^{2} \rangle \tau/(\gamma_{\mu}^{2} 
B_{\mathrm{a}}^{2} \tau^{2}+1),
$
which suggests that we are in the limit that $\gamma_{\mu}^{2} B_{\mathrm{a}}^{2} \tau^{2} \ll 1$, 
or, given that the maximum field was 300~mT, $\tau \lesssim 5$~ns.
The main result of these measurements is that, in both HTI and HTII phases, 
the presence of the electric field $E$ results in a reduction of the relaxation rate 
$\lambda(E)$ compared to $\lambda(E=0)$ of $\sim 0.5$~MHz. 
Application of the paired-samples t-test shows that the E-field causes a statistically significant
decrease at a $>99$\% confidence level in both phases. 
Fig.~\ref{efield}(b) and (d) show the normalized difference in the relaxation rates
with the field on and off, (i.e.\ $[\lambda(E=0)-\lambda(E)]/ \lambda(E=0)$ as a function of
$B_{\mathrm{a}}$. Taking the average of these results across the B-field range
shows the same average decrease in this quantity of 11(1)\% for each phase. 

In interpreting these E-field dependent data we note first that the effect of applied E-field
is the same within experimental uncertainties 
in the HTI and HTII phases. This is in keeping with the optics 
results \cite{lottermoser} where the ME effect was seen at all temperatures below $T_{\mathrm{N}}$ and would imply
that the change in Ho ordering at $T_{\mathrm{SR}}$ is not a prerequisite for the observation of a ME effect. 
Noting that in the fast fluctuation limit the relaxation rate $\lambda \sim \langle B^{2} \rangle \tau$,
it is likely that the decrease in $\lambda$ caused by the application of the E-field reflects a decrease in the
quantity $\sqrt{\langle B^{2} \rangle}$, the local magnetic field distribution at the muon sites. 
(It seems unlikely
that the E-field would cause a  decrease in the correlation time $\tau$.)
There are two ways to decrease $\sqrt{\langle B^{2} \rangle}$: 
(i) a reduction in the magnitude of the 
average local magnetic field at the muon sites, caused, 
for example,
by the E-field altering the magnetic moment structure via the magnetoelectric effect 
in such a way that
the average dipole field at the muon sites decreases;
(ii) a decrease in the width of the magnetic field
distribution. For the case (ii) the E-field leads to an effective increase in the order in the magnetic structure. 
This, we believe, provides the most likely mechanism for our measurements. 

The primary effect of applying a large E-field to a FE is  to produce a single ferroelectric domain. 
Furthermore, it has been demonstrated experimentally in the case of YMnO$_{3}$ \cite{fiebig02} 
(and proposed for HoMnO$_{3}$ \cite{lorentz04}) that FE domain walls coincide with AFM domain
walls. This may be explained by supposing that the lattice strain, which is 
present by virtue of the existence of the
FE domain wall, couples to local Mn magnetic moments. This means that the free energy of the 
system is lowered when
the magnetic order parameter (i.e.\ the sublattice magnetic moment in an AFM) changes sign 
across a FE domain wall, creating a spacially coincident AFM wall. 
This implies that all FE domain walls coincide with
AFM domain walls, but that AFM domain walls may exist within a single FE domain. The application of the
E-field therefore causes not only a single FE domain, but also a reduction in the number of AFM domains.
This will reduce the width of the magnetic field distribution 
$sqrt{\langle B^{2} \rangle}$
probed by the muons and causes the small observed
decrease in the relaxation
rate $\lambda$.
Our E-field $\mu^{+}$SR results may therefore be consistent with the magnetoelectric
coupling in HoMnO$_{3}$ being mediated via domain walls, as was originally proposed to explain the
results of dielectric constant measurements \cite{lorentz04}.

\section{Conclusion}

In conclusion, we have studied the ZF and E-field $\mu^{+}$SR of HoMnO$_{3}$. 
We are able to confirm the presence of
magnetic transitions at $T\approx 76$~K, $T=34$~K and $T=38$~K. 
The muon probe is primarily sensitive to the ordering and dynamics of the 
Mn$^{3+}$ magnetic moments, with the Ho ordering having little effect on our measurements. 
The ordering of the Ho moments, at most, contributes via a minority relaxation channel.
The application of electric fields of $E=10^{4}$~Vm$^{-1}$ causes only a $\sim 10$\% change in
the relaxation rate, which may be accounted for by a reduction in the number of
 coupled FE and AFM domain walls. 

\acknowledgments 

Part of this work was carried out at the ISIS Facility, Rutherford Appleton Laboratory, UK 
and at the Swiss Muon Source, 
Paul Scherrer Institute, Villigen, Switzerland. 
We thank A.\ Amato for technical assistance and W.\ Hayes for useful discussions.
This work is supported by the EPSRC.

\end{document}